\documentstyle[epsfig,bbm,12pt,amsmath,amssymb]{article}

\input{epsf.sty}

\def\be{\begin{equation}}
\def\ee{\end{equation}}
\def\baray{\begin{eqnarray}}
\def\earay{\end{eqnarray}}
\def\ba{\begin{eqnarray}}
\def\ea{\end{eqnarray}}

\def\2pi{\left(2\pi\right)}

\def\hf{\frac{1}{2}}
\def\ap{\alpha^{\prime}}

\begin{document}
\bibliographystyle{plain}
\title{Cosmic String Lensing and Closed Time-like Curves}

\author{Benjamin Shlaer\footnote{Electronic mail: shlaer@lepp.cornell.edu}\,
and S.-H.~Henry Tye\footnote{Electronic mail: tye@lepp.cornell.edu}}

\vspace{1mm}

\maketitle

\begin{center}
Laboratory for Elementary Particle Physics\\ 
Cornell University, Ithaca, NY 14853\\
\end{center}

\medskip

%\date{\today}

\abstract{In an analysis of the gravitational lensing by two relativistic cosmic strings, 
we argue that the formation of closed time-like curves proposed by Gott is unstable in 
the presence of particles (e.g. the cosmic microwave background radiation). Due to the 
attractor-like behavior of the closed time-like curve, we argue that this instability is very 
generic. A single graviton or photon in the vicinity, no matter how soft, is 
sufficient to bend the strings and prevent the formation of closed time-like curves.
We also show that the gravitational lensing due to a moving cosmic 
string is enhanced by its motion, not suppressed.}

%\maketitle

\section{Introduction}

Although cosmic strings as the seed of structure formation\cite{VilenkinS} has been ruled 
out by observations, their presence at a lower level is still possible. Indeed,
cosmic strings are generically present in brane inflation in superstring theory, and 
their properties are close to, but within all observational bounds  
\cite{Jones:2002cv,
%Sarangi:2002yt,Jones:2003da,
Pogosian:2003mz,Kachru:2003sx,Copeland:2003bj,
Kibble:2004hq,Firouzjahi:2005dh}.  
This is exciting for many reasons since current and near
future cosmological experiments/observations will be able to confirm or 
rule out this explicit stringy prediction. If detected, the rich properties 
of cosmic strings as well as their inflationary signatures provide a window to both 
the superstring theory and our pre-inflationary universe. 
The cosmic strings are expected to evolve to a scaling string network with a spectrum of tensions. Roughly speaking, the physics and the cosmological implications are entirely dictated by the ground 
state cosmic string tension $\mu$, or the dimensionless number $G\mu$,
where $G$ is the Newton's constant \cite{VilenkinS}.
The present observational bound is around $G\mu \lesssim 6 \times 10^{-7}$.
Recently, it was shown \cite{Firouzjahi:2005dh} that the cosmic string tension can easily 
saturate this bound in the simplest scenario in string theory, namely, the realistic 
$D3-\bar D3$-brane inflationary scenario \cite{Kachru:2003sx}.
This means gravitational lensing by such cosmic strings, providing image separation 
of order of an arc second or less, is an excellent way to search for cosmic strings. 
Generic features of cosmic strings include a conical ``deficit angle'' geometry, so a
straight string provides the very distinct signature of an undistorted double image.
Cosmic string lensing has been extensively studied \cite{Vilenkin:1984ea,
Vilenkin:1984ib,Vilenkin:1986cb,deLaix:1996vc}.
%We shall briefly review the lensing by straight moving cosmic string. 

In the string network, some segments of cosmic strings will move at relativistic speeds.
It is therefore reasonable to consider the gravitational lensing by highly relativistic cosmic strings.
We also confront another interesting feature of cosmic strings, first realized by 
Gott \cite{Gott:1990zr}:
the possible appearance of closed time-like curves (CTCs) from two parallel cosmic strings
moving relativistically past each other. 
As the strings approach each other fast enough in Minkowski spacetime, the path encircling 
the strings in the sense opposite to their motion becomes a CTC. 
This is sometimes called the Gott spacetime or Gott time machine.

Although there is no proof that a time machine
cannot exist in our world \cite{Greene}, 
their puzzling causal nature leads many physicists to believe that 
CTCs cannot be formed. 
This skepticism has been encoded in  
%There is much controversy surrounding CTCs, in particular due to 
Hawking's Chronology Protection Conjecture (CPC) \cite{Hawking:1991nk}.
However, CPC as proposed is not very precise; even if we assume
CPC is correct, it is not clear exactly what law of physics will prevent the specific 
Gott spacetime. There are a number of interesting and insightful studies
attempting to apply CPC against the Gott's spacetime :  

\begin{itemize}

\item 
Recall that the original Chronology Protection Conjecture \cite{Hawking:1991nk} 
is motivated by two results.  The more general of the two is the semi-classical 
divergence of the renormalized
stress-energy tensor near the ``chronology horizon,'' or Cauchy surface separating the 
regions of spacetime
containing CTCs from those without.  These (vacuum)``polarized'' hypersurfaces led Hawking to
conjecture that {\em any} classical spacetime containing a chronology horizon will be excluded from
the quantum theory of gravity.
However, a recent paper \cite{Johnson:2004zq} found an example where this 
chronology horizon is well defined in the background of 
a string theory (to all orders in $\ap$).
So, in superstring theory, the CPC is not generically true: the Taub-NUT geometry receives
corrections that preserve the travsersibility of the chronology horizon.  Unlike classical 
Taub-NUT, stringy Taub-Nut contains timelike
singularities, although they are far from the regions containing CTCs.  The second hint that
CPC is true is the theorem proved by Hawking and Tipler\cite{Hawking:1991nk,Tipler} that spacetimes 
obeying the weak energy 
condition with regular initial data and whose
chronology horizon is compactly generated cannot exist.  These theorem depends crucially on
the absence of a singularity, and so Hawking's claim that finite lengths of cosmic string
cannot produce CTCs is only true if one rejects the possibility of a singularity being present somewhere
on the chronology horizon\cite{Carroll:1994hz}.

\item When we consider possible formation of CTCs coming from cosmic string
loops (though
this is not necessary for our general argument), Tipler and Hawking make use of the null
energy condition and smoothness
to argue against CTCs. The null energy condition
can be satisfied even when
one smooths out the conical singularity (with a field theory model) at
the core of the
cosmic strings. However, in superstring theory, such a procedure is not
permitted.
The cosmic strings are either D1-strings or fundamental strings.
Classically, the core
is a $\delta$-function with no internal structure (e.g., energy
distribution) in the string
cross-sections, so the string has only transverse excitation modes.
Suppose we smooth out the $\delta$-function. Then one can rotate the
string around its axis and endow it with longitudinal modes. The
presence of such longitudinal modes violate the unitarity property of
the superstring theory. (In fact, in the presence of such longitudinal
modes, general relativity is no longer assured in superstring theory.) 
In this sense, the string geometry is not differentiable, and one must 
generalize the appropriate theorems before they may be applied.

\item Cutler \cite{Cutler} has shown that the Gott spacetime contains regions free of CTCs, 
that the chronology horizon is classically well defined, and that Gott spacetime contains no
closed (as opposed to just self-intersecting) null geodesics.  Hawking points out that this
last feature must be discarded for bounded versions of Gott (and similar) spacetimes.  This means
that if one can avoid the Tipler and Hawking theorems (by including singularities), a cosmic string
loop could create a local version of Gott spacetime {\em with} closed null-geodesics.
Cutler found a global picture
of the Gott spacetime very much in agreement with general arguments made by Hawking regarding 
the instability of Cauchy horizons, 
specifically
the blue-shifting of particles in CTCs.  Here, we find a concrete example of this phenomenon using
a lensing perspective.

 \item  As parallel strings move relativistically past each other to create CTCs, a black hole 
 may be formed by the strings before the CTC appears, thus preventing CTCs. If this happens, 
 one can consider the formation of a black hole as a realization of CPC. However, it is easy to 
 see that, using Thorne's hoop conjecture, there is a range of string speed where the CTCs 
 appear, but no black hole is formed.  The results of Tipler and Hawking suggest that either the strings
 are slowed to prevent CTCs, or a singularity forms somewhere else in the geometry.   We will argue that the strings
 are slowed, and no CTC forms.

 \item Tipler\cite{Tipler} proved that whenever a CTC is produced in a finite region of spacetime, 
a singularity must necessarily accompany the CTC.  This singularity does not represent a no-go
theorem, since the CTC and its sources need not encounter the singularity.  In fact Tipler's physical
argument against the creation of CTCs is the unfeasibility of creating singularities. 
However, it is well-known that singularities such as orbifold fixed points and conifolds 
are perfectly fine in superstring theory, where Einstein gravity is recovered as a low energy 
effective theory. Furthermore, under the appropriate circumstances,  topology changes are 
perfectly sensible. This consistency is due to the extended nature of string modes.

 \item One may consider the Gott spacetime in 2+1 dimensions. The 2+1 dimensional gravity 
relevant for the problem has been studied by Deser, Jackiw and 't Hooft \cite{Deser:1983tn}. 
For a closed universe, 't Hooft \cite{tHooft2} argues that the universe will shrink to zero volume
before any CTCs can be formed.  For an open universe, Carroll, Farhi, Guth and 
Olum (CFGO) \cite{Carroll:1994hz} show that it will take infinite 
energy to reach Gott's two-particle system which has spacelike total momentum. 
However, the argument depends crucially on the 
dimensionality of spacetime. We argue that this last property is quite 
specific to 2+1 dimensions. In 3+1 dimensions, we show that it is easy to realize Gott's
two-string system. For example, a long elliptical string with slowly moving sides will 
collapse to two nearly parallel segments at high velocity, and can do so
without forming a black hole. So CTC formation from the evolution of cosmic 
string loops seems quite easy to construct. This feature is purely 3+1 dimensional.
 
\end{itemize}

If none of the above arguments against CTCs are fully applicable to the Gott 3+1 spacetime,
does this mean CPC fails and Gott spacetime can be realized in the real universe?
Or are there other mechanisms which prevent the formation of CTCs?
  
In this paper, we use a lensing framework to demonstrate the classical instability near the
Cauchy horizon
which we argue will prevent the
formation of cosmic string CTCs in any realistic situation. To be specific, our argument 
goes as follows :

\begin{itemize}

\item A particle or a photon gets a positive kick in its momentum in the plane orthogonal 
to the strings each time it goes around a CTC \cite{Kaiser:1984iv,Carroll:1994hz,Carroll:1992gc}.

\item Once inside the chronology horizon, such a particle is generically attracted to a CTC; 
that is, a worldline in the vicinity of the CTC will coalesce with the CTC. This is 
our main observation.

\item The particle will go through the CTC numerous times (actually an infinite number 
of times) instantaneously; that is, the particle will be instantaneously 
infinitely blue-shifted.

\item It follows that the back-reaction must be important; conservation of angular 
momentum and energy implies that the cosmic strings will slow down, or, more 
likely, bend; this in turn prevents the formation of CTCs.  Note that this back 
reaction must disrupt the closed time-like curve, 
otherwise the  infinite blue-shift can not be prevented.  Thus a single particle, say a graviton 
or photon, no matter how soft, will bend the cosmic strings so that CTC cannot be formed.
The following picture seems reasonable: as the two segments of cosmic strings move
toward each other, they are bent and so radiate gravitationally. This slows them down 
to below the critical value for CTC formation.
We expect no singularity/divergence to appear. 

\item Since there is a cosmic microwave background radiation in our universe, 
these photons preclude the existence of CTCs. Of course, the cosmic microwave background 
radiation is not the only wrench in the machine. 
Gravitons or some other particles can be emitted by the moving strings, either classically or via 
quantum fluctuation. In particular, gravitons must be present in spacetimes of dimensions 3+1 
or greater. A single graviton, no matter how soft, will lead to the above effect. We argue this is 
how the chronology protection conjecture works in the Gott spacetime.

\end{itemize}

This result is not too surprising in light of the likely (blue-shift) instability of Cauchy horizons discussed
by Hawking and others, although a counter example was found by Li and Gott \cite{Li:1997ka} while analyzing possible vacua of
Misner space, whose Cauchy horizon can be free of instability.  As in our example, a divergence occurs only
in the presence of particles, although we find that blue-shifting (and not particle number) is the cause.

The blue-shift instability is well studied in the literature \cite{Poisson:1997my}.
The strong cosmic censorship conjecture predicts that a Cauchy horizon is, in general unstable 
(e.g. that of a Reisner-Nordstr\"om
black hole in asymptotically flat spacetime), and that this instability is the
result of the infinite blue shift of in falling perturbations.  This must be similar to
the instability we describe, but our picture is resolved differently.  We find that
the CTC never forms because surrounding particles scatter off the cosmic strings, bending and slowing them.
Hence no Cauchy horizon (stable or not) ever forms.

't Hooft\cite{tHooft3} argues that, since the local equations of motion for a cosmic string 
are well-defined, one should be able to list the Cauchy data at any particular time, and
demand the Laws of Nature to be applied in a strictly causal order. 
If one phrases the logic this way, there are no CTCs by construction, in agreement 
with the chronology protection conjecture and strong cosmic censorship. So the question is:
what is wrong with Gott spacetime?  His answer to this question is
that the Cauchy planes become unstable: in terms of these, the Universe
shrinks to a line in 3+1 dimensions. The moment a disturbance from
any tiny particle is added somewhere in the past, it generates so much
curvature that the inhabitants of this universe are killed by it.
In our scenario, we give a specific mechanism with more details: 
a single graviton or photon, even a very soft one, will suffice. 
An infinitely blue-shifted photon (or any particle) will cause so 
much curvature that 't Hooft's collapsing scenario occurs. Here, we 
agree with the chronology protection principle and 't Hooft that a 
CTC is not formed. However, we believe that, due to the
energy-momentum-angular momentum conservations, the back-reaction
will bend the cosmic strings and induce gravitational radiation so that the CTC is never formed.
Neither the curvature nor the energy of the photon blows up. Note that the bending of 
the strings can not happen in 2+1 dimensions. 

In this paper, the motion of a photon/graviton around the cosmic strings is a crucial ingredient
of the analysis. We shall start with a review of the gravitational lensing by a straight 
moving cosmic string. Here we correct a mistake lensing formula mistake in the literature (see Appendix A). 
Next we review the evolution of a simple string loop to a loop with two long segments
that are moving past each other at ultra-relativistic speed. Far away from the ends, we 
treat the two long segments as if they were two infinite parallel strings. Next we review 
Gott spacetime. Finally we show that the CTCs encircling the two strings are attractors
for particles. Our analysis ends here. Supplemented with plausible reasonings,
we argue that the above mechanism is a way to prevent CTCs in the real universe.
Throughout, we shall assume $\delta_0 =8 \pi G \mu$ to be very small (say, less than $10^{-5}$).

\section{Cosmic String Lensing}

One may calculate the observational signatures of rapidly moving cosmic strings
(straight and loops), in particular their lensing effects on distant galaxies
and the CMB. The simplest case of a straight, nearly static cosmic string has the
distinctive signature of producing two identical images, each being undistorted
and equidistant from the observer.

\begin{center}
\epsfig{file=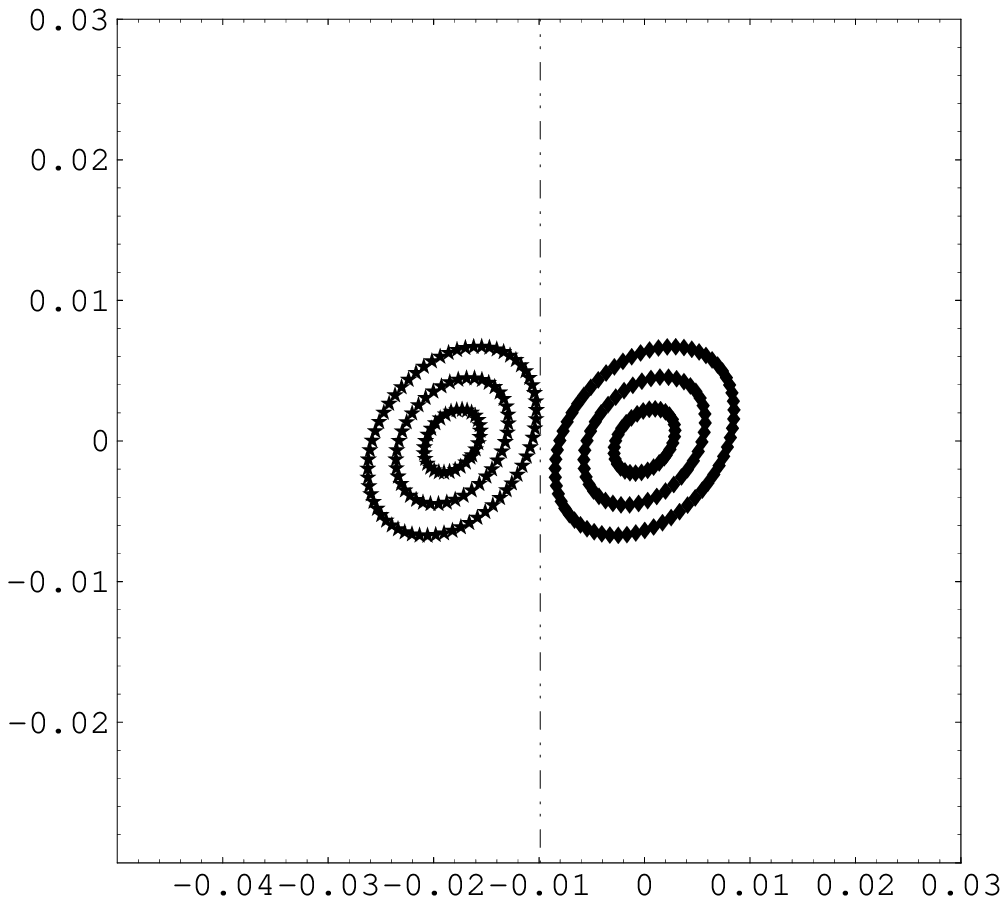, width=6cm}\epsfig{file=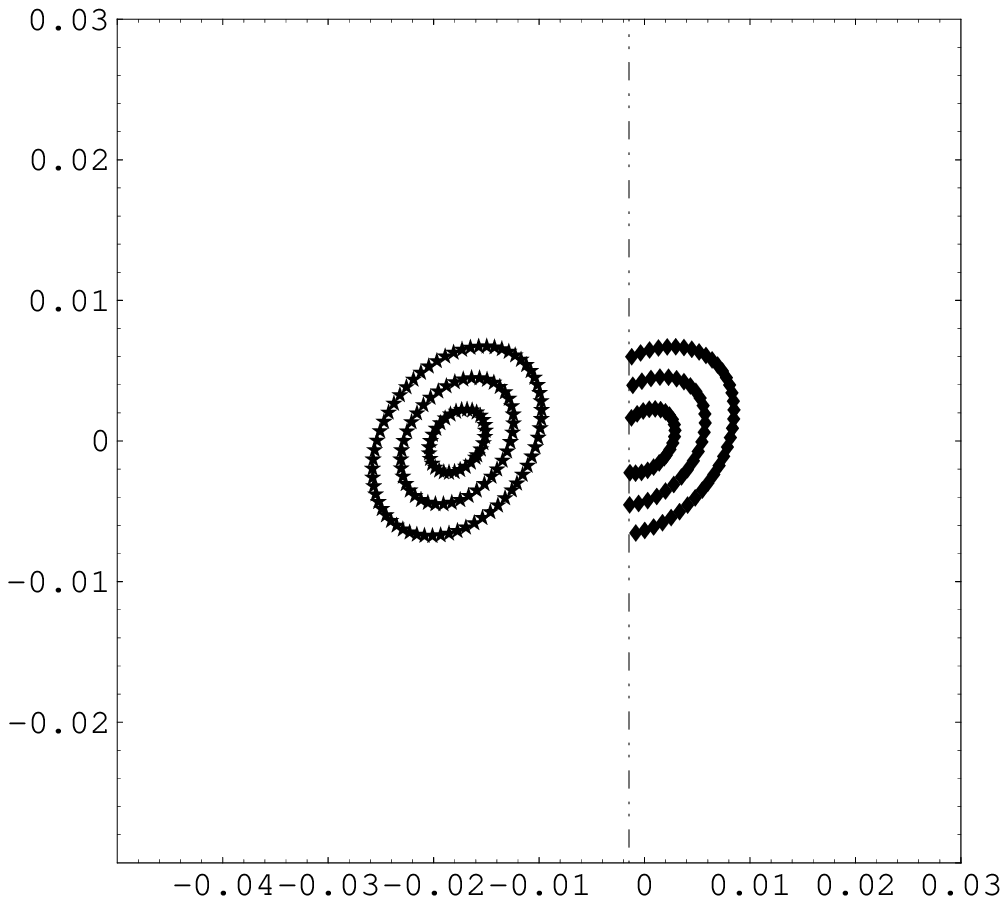, width=6cm}\\
\end{center}
\noindent {\bf FIG 1}. Gravitational lensing by a straight cosmic string. 
The two images are identical. Note that they are not mirror
reflections
of each other, but one is a translational displacement of the other.
As the source moves to the left (or equivalently, the cosmic string
moves to the right), part of the right image is cut off, eventually leaving only a
single image.

\vspace{4mm}

Above is pictured a cosmic string moving to the right across a distant galaxy.
We call the angular separation of the images
$\delta\varphi$, and the photon deflection angle $\delta$, as in the
diagram below. Although the double images on the left picture of Fig. 1 may be 
due to two almost identical galaxies (a rare but not impossible scenario), the picture 
on the right will be a much cleaner signature of cosmic string lensing. If one sees a 
double image candidate \cite{csl}, one expects to see other candidates 
nearby. Searching for incomplete images will be important.

\begin{center}
\epsfig{file=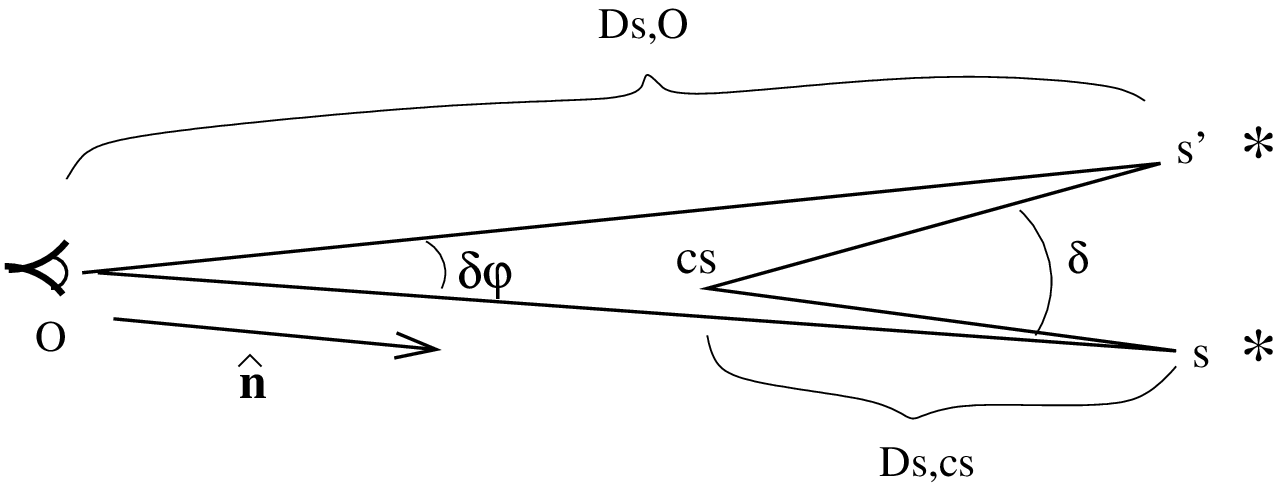 ,width=10cm}\\
\end{center}
\noindent {{\bf FIG 2}. A straight cosmic string introduces a deficit angle $\delta$. Here,
the sources $s$ and $s'$ are to be identified.

\vspace{4mm}

This leads to the well known result $\delta\varphi = \frac{D_{s,cs}}{D_{s,O}}\delta$.
The spacetime around a static cosmic string is Minkowski space with the
identification of two semi-infinite hyperplanes whose intersection is
the cosmic string world sheet.  This is equivalent
to identifying every event $s$ in space time with a dual $s'$ where
the relation between $s$ and $s'$ with a static cosmic string located
at $r_{cs}$ is
\be
s' = R_{\delta_0}(s-r_{cs}) + r_{cs}.
\ee
Here $R_{\delta_0}$ is a pure rotation (counter clockwise).  (Notice $r_{cs}$ 
can be any point on the cosmic string world-sheet.)  It should be noted that $s$ is visible 
only when it appears to the right of the cosmic string, and $s'$ is visible only when
to the left.

The general case involves a cosmic string moving at some four-velocity $v_{cs}$. 
We will always take $v_{cs}$ to be perpendicular to the cosmic string world
sheet, since any parallel component is unphysical (assuming a pure tension string).
Interestingly, this velocity is only well defined in combination with 
a ``branch cut''. This is related to the fact that a passing cosmic string will 
induce a relative velocity
between originally static points in space, so a constant velocity field will not be everywhere 
single valued.  More physically, parallel geodesics moving past a cosmic string will be bent toward
each other, provided they pass the string on opposite sides.  Specifying a branch
cut enables the conical geometry to be mapped to Minkowski space (minus a
wedge), where things are simpler. 
The pure rotation identification is 
valid only in the center of mass frame of the cosmic string, so in general
\be
s' = \Lambda_{v_{cs}}R_{\delta_0}\Lambda_{-v_{cs}}(s-r_{cs}) + r_{cs}
\ee
where $\Lambda_{v_{cs}}$ is a pure boost such that $\Lambda_{-v_{cs}}v_{cs} = 
(1,0,0,0)$.  We have simply boosted into the strings reference
frame and then performed the rotation-identification.  Then we boost back.

\begin{center}
\epsfig{file=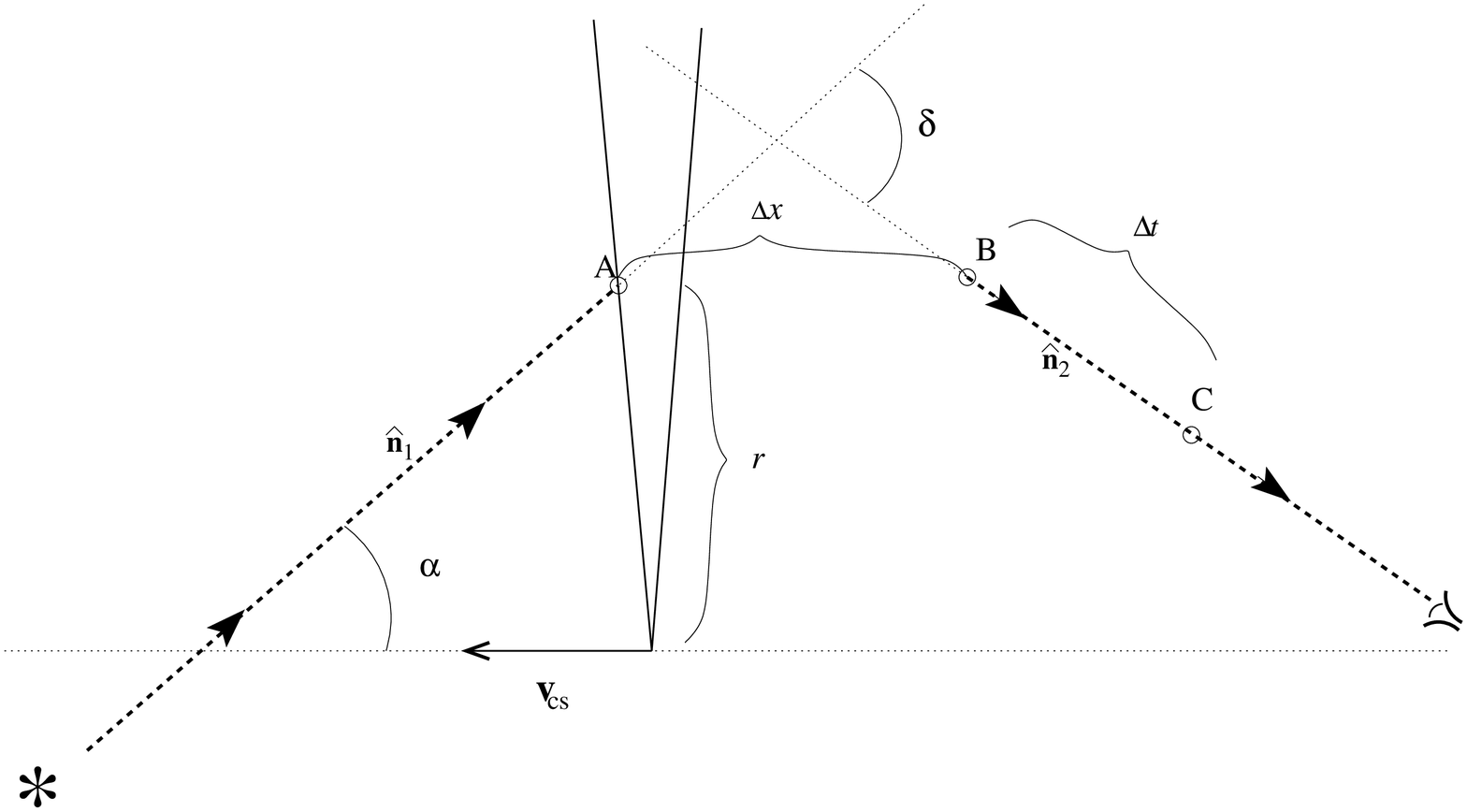 ,width=11cm}\\
\end{center}
\noindent {{\bf FIG 3}. The path of a photon coming from the source at the lower left 
and reaching the observer at the right. 
The straight cosmic string is moving with speed $v_{cs}$ to the left. The 
photon's initial velocity $\hat{{\bf n}}_1$ makes an angle $\alpha$ 
with respect to $- {\bf v}_{cs}$. At A the photon strikes the leading 
edge of the deficit wedge (a distance $r$ from the 
position of the cosmic string).

\vspace{4mm}

Here we investigate the lensing due to nearly straight segments moving at arbitrarily
relativistic speeds.  We consider the
interaction of a photon with a cosmic string's deficit angle.  In Figure 3, a photon 
is crossing the deficit angle at a distance $r$ from the string vertex.
The photon is re-directed by an angle $\delta$ and makes a spacial and temporal
jump.  This jump  ($s' - s$)  is found using the coordinate identification in Eq. (2), where $s = $A is the 
photon striking the deficit angle and $s' = $B is its emergence from the deficit angle.  If we choose the
deficit angle to be perpendicular to $\bf{v}_{cs}$, the spacial jump is parallel to $\bf{v}_{cs}$,
and is  given by
\be
\Delta x = 2r\tan(\delta_0/2)\gamma_{cs}\quad\quad\quad \Delta t = -2r\tan(\delta_0/2) \gamma_{cs} v_{cs}.
\label{xt3}
\ee
where  $\gamma_{cs} = 1/\sqrt{1 - v^2_{cs}}$.
In Figure 3, events A and B are identified, while events A and C are simultaneous.
If the photon strikes the leading edge of the deficit angle, the jump is behind the cosmic string
and backward in time.  If the photon strikes the trailing edge, the jump is in front of the cosmic
string and forward in time.  For ultra relativistic cosmic strings, only photons traveling
almost exactly parallel to the string's velocity will strike the trailing edge of the deficit angle.

We will calculate the change in momentum of a particle interacting with the cosmic string.  
In the string rest frame, we know that 
\be
k_{final} = R_{\delta_0} k_{initial}
\ee
so we simply boost the above equation into the frame where the string is moving at 
velocity $v_{cs}$.  Then
\be
k_{final} = \Lambda_{v_{cs}}R_{\delta_0} \Lambda_{-v_{cs}}k_{initial}.
\ee

To calculate the directional change in a photon's velocity, we take the above formula with
$k^2 = -m^2 = 0$ and for simplicity we take the photon to travel in a plane perpendicular 
to the cosmic string.  Then we find (dropping the $cs$ subscript)
\baray
\label{deltaalp}
&&\cos(\delta) = \\
&&\frac{((A - B)\cos(\alpha)
+\sin(\alpha)(v\sin(\delta_0)-\cos(\delta_0)\sin(\alpha)/\gamma))}
{\sqrt{(\sin(\delta_0)(v + \cos(\alpha))-\cos(\delta_0)\sin(\alpha)/\gamma)^2 + 
(B - A + \sin(\alpha)\sin(\delta_0))^2}}  \nonumber
\earay
where
\be
A = \gamma\cos(\alpha)(v^2-\cos(\delta_0))\quad\quad\quad B=\gamma(v\cos(\delta_0)-1).\nonumber
\ee

The blue-shift can be calculated as well, yielding
\be
\frac{\omega_{final}}{\omega_{initial}} = \gamma \sin(\alpha)\sin(\delta_0) + \gamma^2(1 + v\cos(\alpha) - 
v\cos(\alpha)\cos(\delta_0)-v^2\cos(\delta_0)).
\label{blueshift}
\ee
Straight portions of a cosmic string moving at ultra relativistic speeds produce
photon deflection of order $\pi$ in conjunction with severe blue-shift.  Slower moving cosmic strings
will obey the Kaiser-Stebbins formula\cite{Kaiser:1984iv} and cause the sky behind the moving string to
be blue shifted relative to the sky in front of the string.

With the exception of loops, we expect cosmic strings to move
moderately relativistically, but with $\gamma\delta_0 << 1$.  In this limit, the above formula
reduces to
\baray
\delta &=& \delta_0 \gamma (1 + v\cos(\alpha))  \quad \to   \nonumber \\
\delta\varphi &=& \frac{D_{s,cs}}{D_{s,O}}\times\frac{8\pi G\mu}{\sqrt{1-v^2}}
\times (1 + \hat{{\bf n}}\cdot {\bf v}).
\earay
The first factor ${D_{s, cs}}/{D_{s,O}}$ is a plane-geometric coefficient 
for $8\pi G\mu\gamma$,
which is the relativistic energy of the string.
The third term is the result of the finite travel time of light, and does not
represent the coordinate locations of the images in the observer's frame. Notice that
a moving string (except one moving toward the observer) has a stronger lensing effect.
This result disagrees with that given in Ref.\cite{Vilenkin:1986cb}, which has the $\gamma$ 
factor in the denominator. To see this difference more clearly, we give in Appendix A the 
simple derivation of Ref.\cite{Vilenkin:1986cb} and point out where the error occurs.
Recall that the typical speed of the cosmic strings in the network is rather large, $v \sim 2/3$ \cite{Wyman}.
There will be segments of strings that have $\gamma >> 1$ and they have the best 
chance to be detected via lensing.

The above formula only applies to cosmic strings perpendicular to the
line of sight. For the most general lensing due to straight cosmic strings, 
see Ref.\cite{ShlaerWyman}.

\section{The Evolution of a Simple String Loop}

Naively, ultra-relativistic straight strings are rather unlikely, and two parallel
ultra-relativistic straight strings passing each other with such a large kinetic energy
density must take some arrangement. It is along this line of reasoning that
CFGO \cite{Carroll:1994hz} argues against the formation of Gott spacetime in 2+1 dimensions. 
In 3+1 dimensions, the situation is much more relaxed.
String loops provide ultra relativistic speeds that long cosmic
strings rarely obtain.  
This is favorable for the possibility of closed time-like curve formation.
Here we demonstrate how a cosmic string loop can evolve to long, nearly parallel segments 
moving ultra-relativistically toward each other 
at arbitrarily small impact parameter but without touching. In accordance with the hoop conjecture,
the loop avoids collapsing to a
black hole, seemingly allowing the formation of a Gott spacetime with CTCs.  Such a spactime
contains closed null geodesics.  We will argue that, unlike in $2+1$ dimensions the gravitational
radiation present is enough to preclude the formation of CTCs.  In agreement with Tipler and Hawking,
the loop must slow down.  

%It is interesting that cosmic string loops do not self intersect, generically\cite{Kibble:1982cb}.

%A simple explicit solution to cosmic string loop trajectories has been developed
%by Kibble and Turok \cite{Kibble:1982cb}.  
The classical string equation of motion is
\be
\ddot{{\bf r}}(\sigma, \tau) - {\bf r}''(\sigma,\tau) = 0
\ee
with the constraint equations $\dot{{\bf r}}\cdot{\bf r}' = 0$ and $\dot{r}^2 + r'^2 = R_0^2$.
We'll see that $2\pi R_0\mu$ is the total energy of the loop.
A dot symbolizes differentiation with respect to $\tau = t/R_0$, and a prime denotes differentiation
with respect to $\sigma$.
The general solution ${\bf r}(\sigma,\tau)$ can be written as a linear combination of 
periodic left- and right-moving waves, 
\be
{\bf r}(\sigma,\tau) = \frac{R_0}{2}[{\bf a}(\sigma - \tau) + 
{\bf b}(\sigma + \tau)] 
\ee
where $(a')^2=(b')^2=1$.  Consider a set of initial data
\be
\label{eq11}
{\bf r}(\sigma, 0)  =  {\bf r}_0(\theta),\quad\quad\quad
\dot{{\bf r}}(\sigma, 0)  =  {\bf v}_0(\theta) R_0
\ee
where the unit circle bijection $\theta(\sigma)$ is used to parametrize the
initial data and ${\bf v}_0\cdot{\bf r}_0 = 0$.  The gauge conditions imply  
%Using the gauge condition $r'^2(\sigma,\tau) + \dot{r}^2(\sigma,\tau) = R^2_0$ 
%we can find the relation between $\theta$ and $\sigma$.
\be
\label{eq12}
\sigma(\theta) = \int_0^{\sigma(\theta)}d\sigma' = 
\frac{1}{R_0}\int_0^\theta\frac{|{\bf r}'_0(\theta')|}{\sqrt{1 -  v_0^2(\theta')}}d\theta'.
\ee
We require $\sigma(2\pi) = 2\pi$.  The inverse $\theta(\sigma)$ may then
be found, as well as the general solution
%Then it is straight forward to show
\be
{\bf r}(\sigma,\tau) = \frac{1}{2}\left[{\bf r}_0(\theta(\sigma + \tau)) + {\bf r}_0(\theta(\sigma - \tau)) 
+ R_0\int_{\sigma - \tau}^{\sigma + \tau}{\bf v}_0(\theta(\sigma')d\sigma' \right].
\ee
We can apply this formula to the Gott initial data, namely two parallel segments
of length $l_0 = \pi R_0\sqrt{1 - v_0^2}=\pi R_0/ \gamma_0$ passing arbitrarily close, each with 
speed $v_0$ in opposite directions:
\baray
{\bf r}_0(\theta) &=& \frac{R_0}{\gamma_0}(0,\triangle(\theta),\epsilon \cos(\theta))\\ \nonumber
{\bf v}_0(\theta) &=& -\frac{\sqrt{\gamma_0^2 - 1}}{\gamma_0}(\triangle'(\theta),0,0)
\earay
with
% $\gamma_0 = 1/\sqrt{1 - v_0^2}$ and 
$\epsilon << 1$ (head on collision of the two string segments is avoided with a non-zero $\epsilon$.) 
Illustrated below are $\triangle(\theta)$ and $\triangle'(\theta)$.
\begin{center}
\epsfig{file=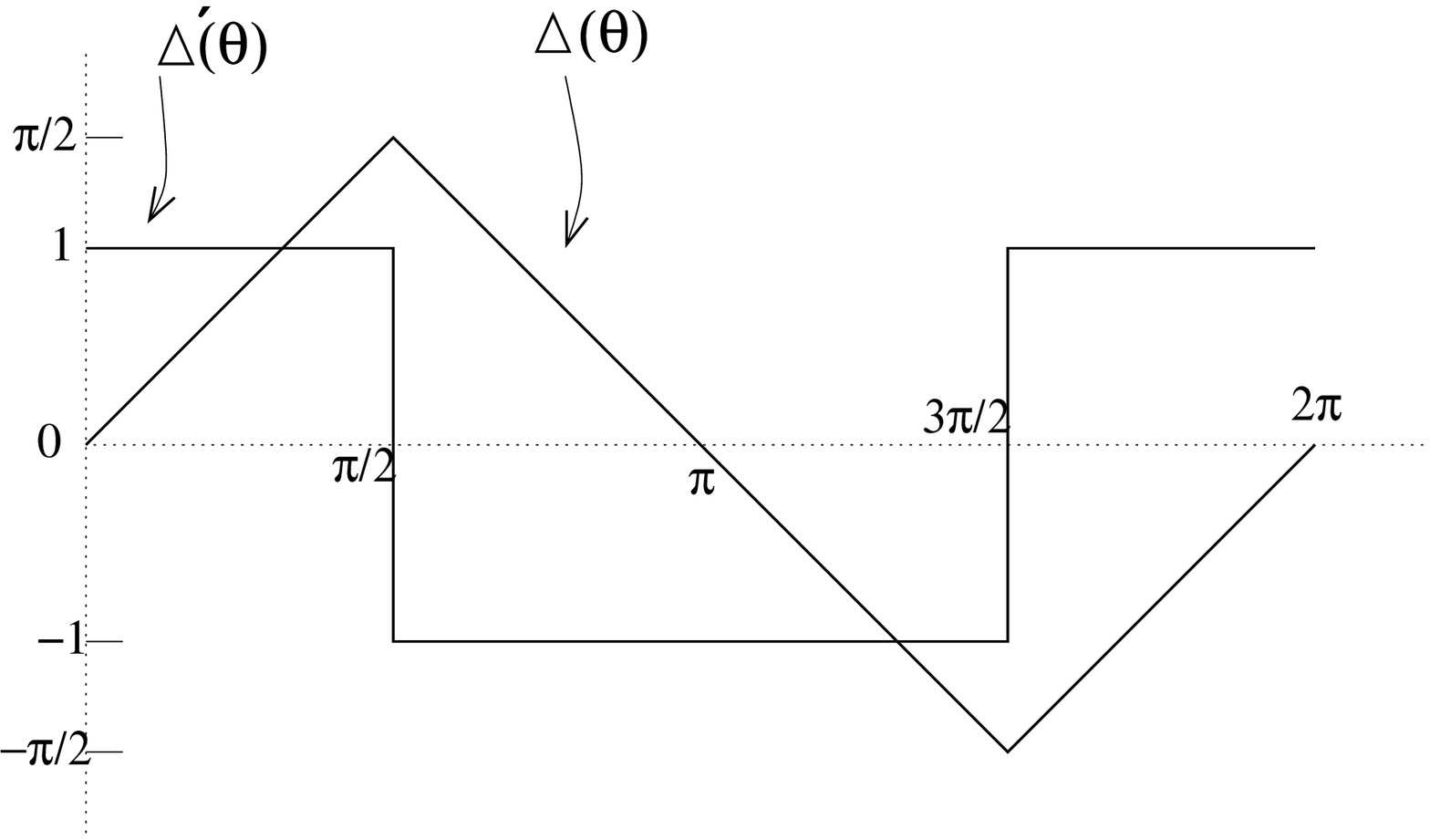 ,width=10cm}\\
\end{center}
\noindent {\bf FIG. 4.}  The triangle function $\triangle(\theta)$ and its derivative $\triangle'(\theta)$.

\vspace{7mm}

Using Eq.(\ref{eq12}) we find $\sigma(\theta) = \theta + O(\epsilon^2)$, and 
\baray
{\bf r}(\sigma, \tau) = \frac{R_0}{2\gamma_0} \lbrack \quad(\sqrt{\gamma_0^2 - 1})[\triangle(\sigma + \tau) &-& 
\triangle(\sigma - \tau)], \nonumber\\
\triangle(\sigma + \tau) &+& \triangle(\sigma-\tau),\\ 
 \epsilon \cos(\sigma+ \tau) &+& \epsilon\cos(\sigma - \tau)\quad \rbrack. \nonumber
\earay

A multi-image snapshot is pictured below.
\begin{center}
\epsfig{file=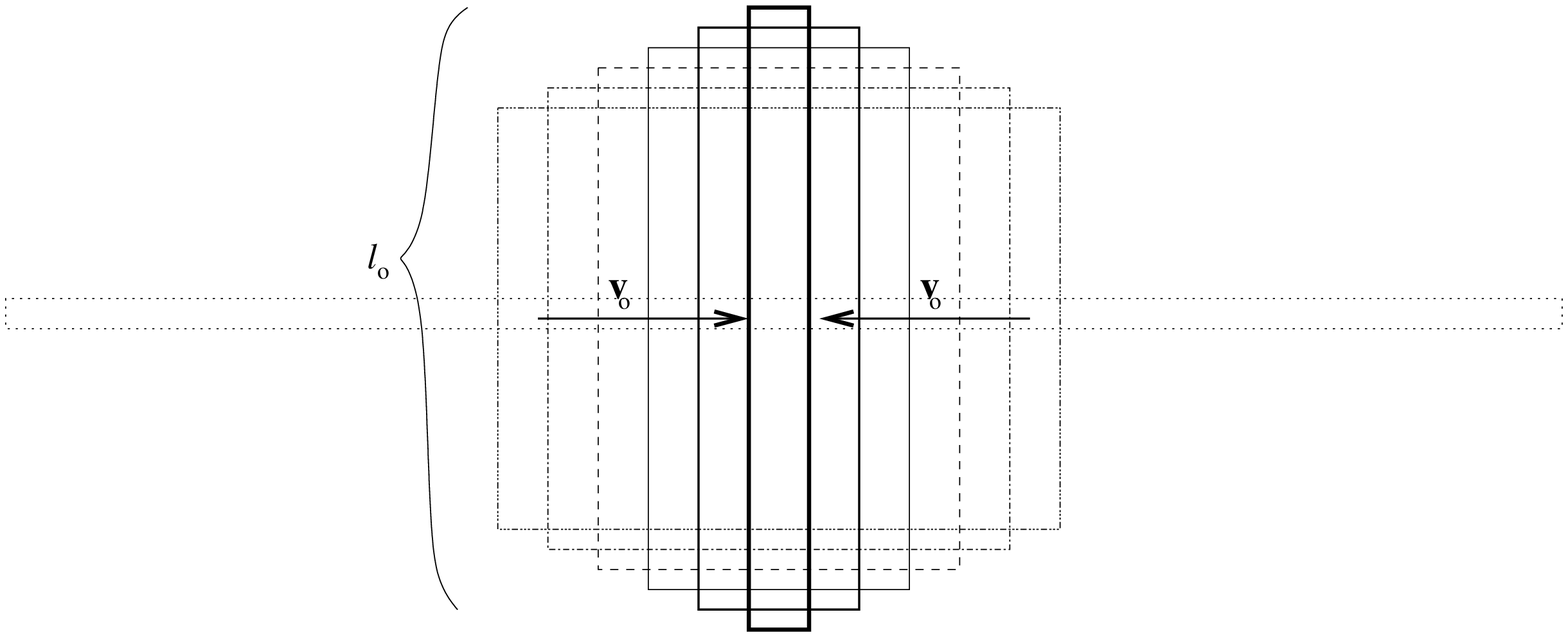, width=13cm}\\
\end{center}
\noindent {\bf FIG. 5.} The Gott-loop has maximum minor and major axes given by $l_0$ and $l_0 v_0 \gamma_0$,
respectively.  The speed of the vertical and horizontal loop sides is given by $v_0$ and $\sqrt{1 - v_0^2}$, 
respectively.

\vspace{7mm}

The first possible obstacle to CTC formation that we consider is Thorne's hoop conjecture.
This states that an event horizon will form (and shroud any CTCs) when and only when a region
of a given circumference $C$ contains more rest energy than the critical energy ${C}/{4\pi G}$. 
We need the loop to collapse to an object with length greater than a critical length $\ell_{crit} =
\delta_0 R_0$.  We immediately see that no  event horizon 
forms for the above solution with $\gamma_0\delta_0 < \pi$.  
This allows for the Gott case $\gamma\delta_0 > 2$.

One can estimate the lifetime of a
cosmic string loop to be of order $100 R_0$ where $R_0$ is the maximum size of
the loop ($2 \pi R_0 = Energy/tension$)\cite{Vilenkin:1984ib}.
Gravitational radiation carries away the otherwise conserved energy
\be
E = \oint \frac{\mu |d{\bf r}|}{\sqrt{1-v^2_\perp}}.
\ee
The time scale of decay is much longer than the time scale of evolution, so one may 
ignore the gravitational radiation. Naively, the inclusion of gravitational radiation may 
only require some minor adjustment of the original string loop.  In reality, the gravitational
radiation plays a role in slowing down the loop sides, preventing CTC formation.

Implicit in this discussion is the assumption that a small region containing parallel segments
of cosmic string loops will have the geometry of infinite strings moving similarly.  This
is a reasonable assumption for the following reason.  Einstein's equations are local and
cannot depend on distant sources (or the lack thereof).  Even if CTCs were to form, they do so in
a bounded region of spacetime (unlike the 2+1 case), and thus can be considered a 
local feature which does not reflect upon distant (spatially or temporally) regions.  One could
imagine adding a small cosmological constant to the 2+1 dimensional Gott spacetime to match
up with an asymptoticly flat universe far from the strings.  
Such a geometry would be sensible for a string loop in
our universe:  Gott spacetime near the loop, yet asymptotically flat.  One would expect that
any process of embedding Gott spacetime in a bounded region would necessarily introduce closed
null geodesics.  In Gott spacetime, null geodesics only close asymptotically as one moves toward
spacelike infinity.

Recent brane inflation models have proposed having the inflationary branes
located in the tip of a Klebanov-Strassler throat of a Calabi-Yau 3-fold \cite{Kachru:2003sx}.
In some scenarios, the standard model branes are in a different throat.  One
feature of this construction is that cosmic strings produced after inflation
are meta stable, and will not be able to decay via open string interaction with
the standard model branes\cite{Copeland:2003bj}.  It should be pointed out that extra 
dimensions have
no effect on Gott's construction of CTCs, provided that the radii of the extra
dimensions are small compared to the radius of the CTC.  
This can be seen as follows.  A cosmic string in our universe is an object
extended in one non-compact direction, and zero or more compact directions.  The
four dimensional effective theory will always have a conical singularity on the
location of the string, and any corrections to this will be due to massive axion and (KK) modes
of the metric-- particles whose range is limited to sizes of order the radii of the
extra dimensions.  Thus as long as we don't probe distances so near the string
that these corrections are significant, the conical geometry is valid and
Gott's construction is meaningful.  In fact, the CTCs in Gott's construction
exist at large radii from the cosmic strings, and so one never needs to
probe the near-string geometry.

Lensing due to entire cosmic string loops has been analyzed
in Ref.\cite{deLaix:1996vc}.

\section{Gott's Construction of CTCs}

Here we give a brief review of Gott's original construction.
The key feature is the conical deficit angle ($\delta_0 = 8\pi G \mu$)
around a cosmic string.  This results in a "cosmic shortcut," since two geodesics 
passing on opposite sides
of the string will differ in length.  This shortcut, like a wormhole, leads
to apparent superluminal travel.  Under boost, ``superluminal'' travel becomes
``instantaneous'' travel.  Gott realized that this ``instantaneous'' travel could be performed in 
one direction, and then back again
when two cosmic strings approach each other at very high speed. The actual 
trajectory is time-like, i.e. performed by a massive body, and resembles an orbit 
around the center of mass of the cosmic string system (in a direction with opposite 
angular momentum as the cosmic strings).  
Below we illustrate the geometry with two strings at rest.

In Figure 6, there are three (geodesic) paths from $A$ to $B$.  The central
path is not necessarily the shortest, in fact it can be seen that
\be
w = x \cos(\frac{\delta_0}{2}) + d \sin(\frac{\delta_0}{2}).
\ee
Thus although $A \to B$ is traversed by a particle on a time-like trajectory above or below
the cosmic strings, the departure and arrival events may have space-like separation
in the $y = 0$ hyperplane which extends between
the two strings.  In this hyperplane, the average velocity of this particle
can be calculated as
\be
v \leq \frac{2x}{2w}\quad = \quad \frac{1}{\cos(\delta_0/2) + \frac{d}{x}\sin(\delta_0/2)}.
\ee
(In our analysis, we focus on light-like motion since time-like motion is, in a
sense bounded by this case.)  
For large enough $x$, this velocity is greater than that of light and so we may boost
to a frame where the departure and arrival events are simultaneous.  This is true for any $d$.
In the above picture, 
we will sever the spacetime at the ($y=0$)-hyperplane and boost such that the top string 
is moving to the left at speed $v_{cs} > 
\cos(\delta_0/2)$ and the bottom string is moving to the right at the same speed.  This
means that we can take $A\to B$ on the upper path and $B\to A$ on the lower path, 
and in both directions the elapsed time is zero.  This is possible when 
\be
2 < 2\gamma \sin(\delta_0/2)\approx \gamma\delta_0.
\ee
It is sufficient that the $y = 0$ hyperplane has vanishing intrinsic and extrinsic
curvature for us to consistently sew the two halves together.  Notice that the limiting
case of $\gamma\delta_0 = 2$ corresponds to closed light-like curves.  (We will assume
that $\frac{d}{x} \to 0$ for simplicity.)

\begin{center}
\epsfig{file=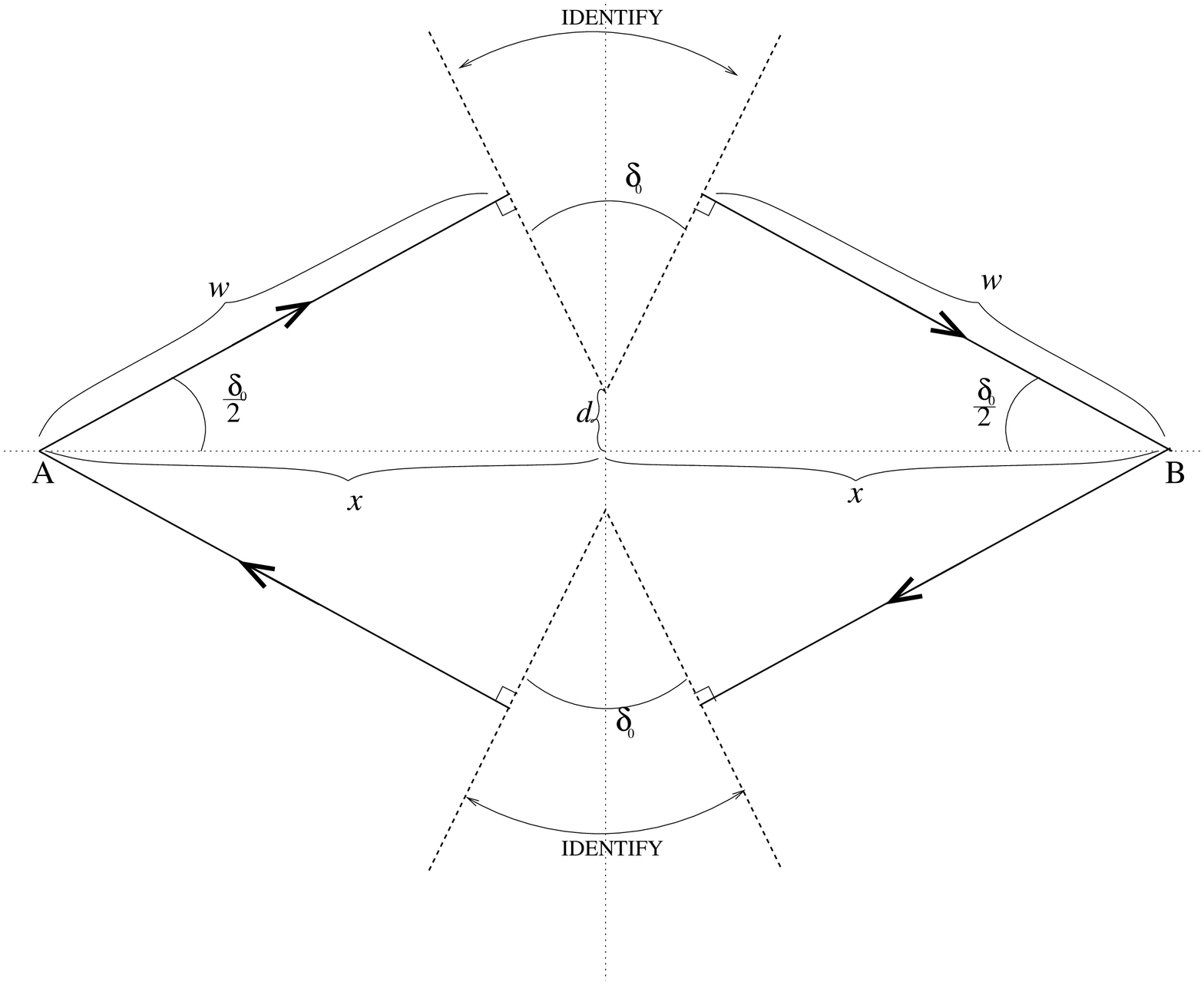 ,width=13cm}\\
\end{center}
\noindent{\bf FIG. 6}. The Gott spacetime, before the strings are moving. This spacetime 
will be severed at the $y = 0$ hyperplane (line AB), boosted, and then 
smoothly glued together again. A and B are separated by a distance of $2x$, while the 
cosmic strings are separated by a distance of $2d$.

\vspace{8mm}

The (two-fold) boosted version of the above setup is pictured below.  The events
are numbered in ``proper'' chronological order (that is, the order in which they occur
on the particle worldline), but the center of mass coordinate frame chronological 
order needs to be explained.  Event
$1$ is the light ray initially traveling up to meet the rapidly moving deficit angle, 
which happens
at event $2$.  This meeting is identified (under Eq.2) with event $3$, although in center-of-mass
coordinates event
$3$ happens before the previous events occur.  Events $1$, $4$, and $7$ are 
cm-simultaneous at $t = 0$ while
events $2$ and $5$ occur at $t_{cm} = +1$, and events $3$ and $6$ at $t_{cm} = -1$.
Notice that $\delta = \pi$.
\begin{center}
\epsfig{file=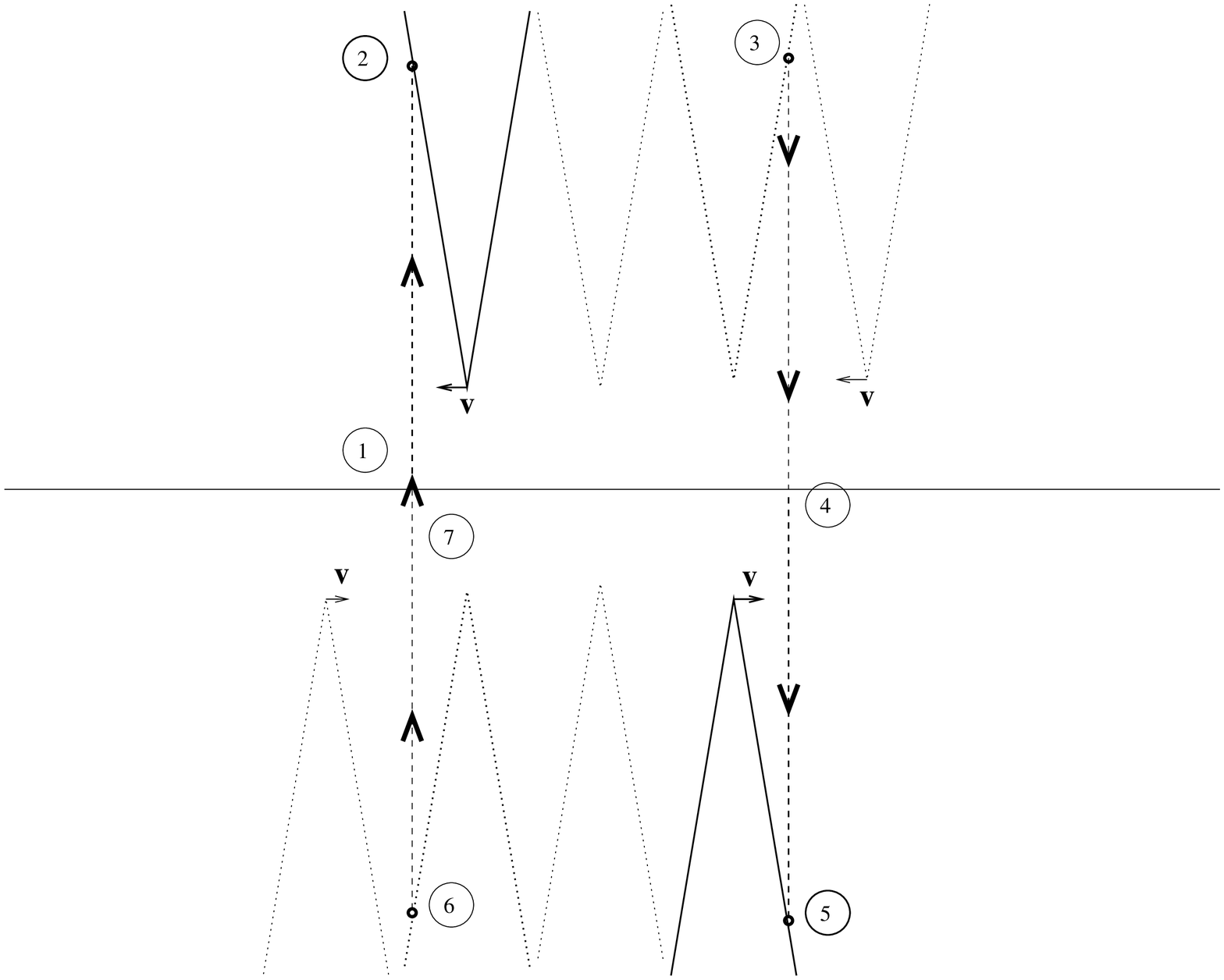 ,width=13cm}\\
\end{center}
\noindent {\bf FIG. 7}. This is the critical case where $\gamma \delta_0 =2$. The deflection angle
is calculated using Eq.(\ref{deltaalp}) and the discontinuity in the world line using Eq.(\ref{xt3}).  
We have drawn $d > 0$ for clarity.  The particle travels
along the path labeled 1 to 7 and back to 1, arriving at the same point in space and time, i.e., a closed 
time-like curve. In this case, the particle is neither blue- nor red-shifted.

\vspace{8mm}

We would like to apply our understanding of ultra relativistic cosmic strings to
Gott spacetime.
We can use the jumps in location, time and direction 
($\Delta t$, $\Delta x$, and $\delta$) to construct all photon paths
around a cosmic string system and determine the complete lensing behavior. 
Below is a graph of $\delta$ for several values of $\gamma\delta_0$, as given by 
Eq.(\ref{deltaalp}).
\begin{center}
\epsfig{file=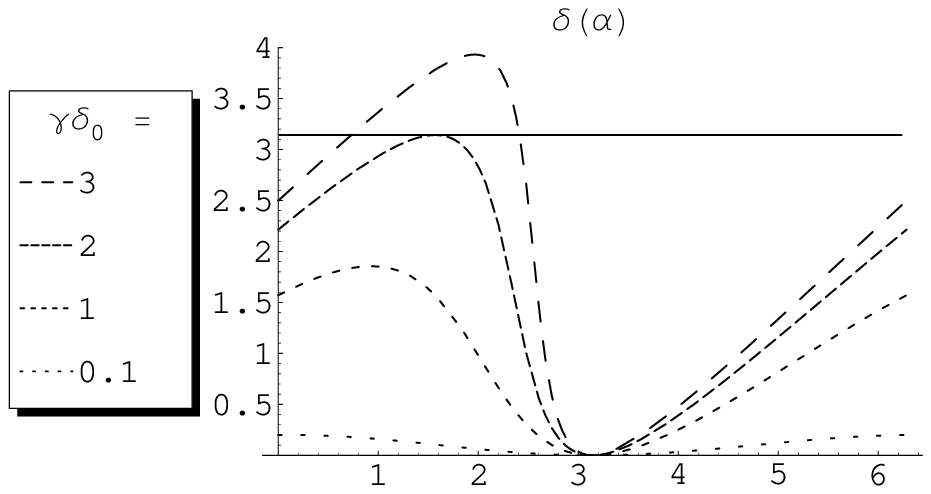,width=14cm}\\
\end{center}
\noindent {\bf FIG. 8}. This plot derives entirely from Eq.(\ref{deltaalp}).  Here, $\delta (\alpha)$ 
is the deflection angle of the photon as a 
function of the angle $\alpha$ between the photon direction and (minus) the
cosmic string velocity (see Figure 3). $\gamma \delta_0=2$ is the critical case, when 
the peak of  $\delta (\alpha)$ touches the value $\pi$ (the solid horizontal line).
A  $\gamma \delta_0>2$ curve crosses  $\delta (\alpha)= \pi$ at 2 points. 
Notable is where $\delta (\alpha)$ crosses $\pi$ (see e.g. the $\gamma \delta_0=3$ curve)
with positive slope at $\alpha_s$. 
A positive slope crossing implies a stable fixed point. Photons with different initial $\alpha$ 
not far from $\alpha_s$ will follow closed time-like paths and approach $\alpha =\alpha_s$. 
Photons are blue-shifted at all positive slope crossings. 
The amount of blue-shift each time is given by Eq.(\ref{blueshift}).

\vspace{4mm}

Of importance is where the graphs take the value $\pi$. Because the two cosmic string 
velocities differ in direction
by $\pi$, $\delta = \pi$ is a fixed point of photon direction.  As Cutler showed, a crossing with 
positive slope is stable and
blue-shifted, while the one with negative slope is unstable and red-shifted.  
The stable blue-shifted fixed point will
always exist for $\gamma\delta_0 > 2$ (super-critical case) and thus represents a 
catastrophic divergence.  This
is because a particle in the presence of this geometry will fall into a stable orbit
with exponentially diverging energy.  The above graph is accurate for massless particles,
but massive particles (equivalently: particles with nonzero momentum along the strings)
will behave similarly once they become blue-shifted. 
In total, we find

\begin{itemize}

\item particle trajectories in the vicinity of a $\gamma\delta_0 > 2$ pair of cosmic strings
will be attracted to the stable orientation $\delta(\alpha) = \pi$, $\delta'(\alpha) > 0$ 
(see Fig. 8).  This is because the two
cosmic strings velocities are equal and opposite, that is, when a particle incident at angle
$\alpha_n$ is deflected by an angle $\delta_n$,  $\alpha_{n+1} = \alpha_n + \pi - \delta_n$.
Thus not only is $\alpha = \pi$ a 'fixed point' of incidence angle, but it is a stable one if
$\delta'(\alpha) > 0$, since then a slight increase in $\alpha$ will cause a slight increase
in $\delta$.  A slight increase in $\delta$ will then decrease the next $\alpha$, and return
the system to equilibrium.  (This assumes $\delta'(\alpha) < 1$, which is always the case.)
Figure 9 illustrates the attractor in action.
\end{itemize}

\vspace{3mm}

\begin{center}
\epsfig{file=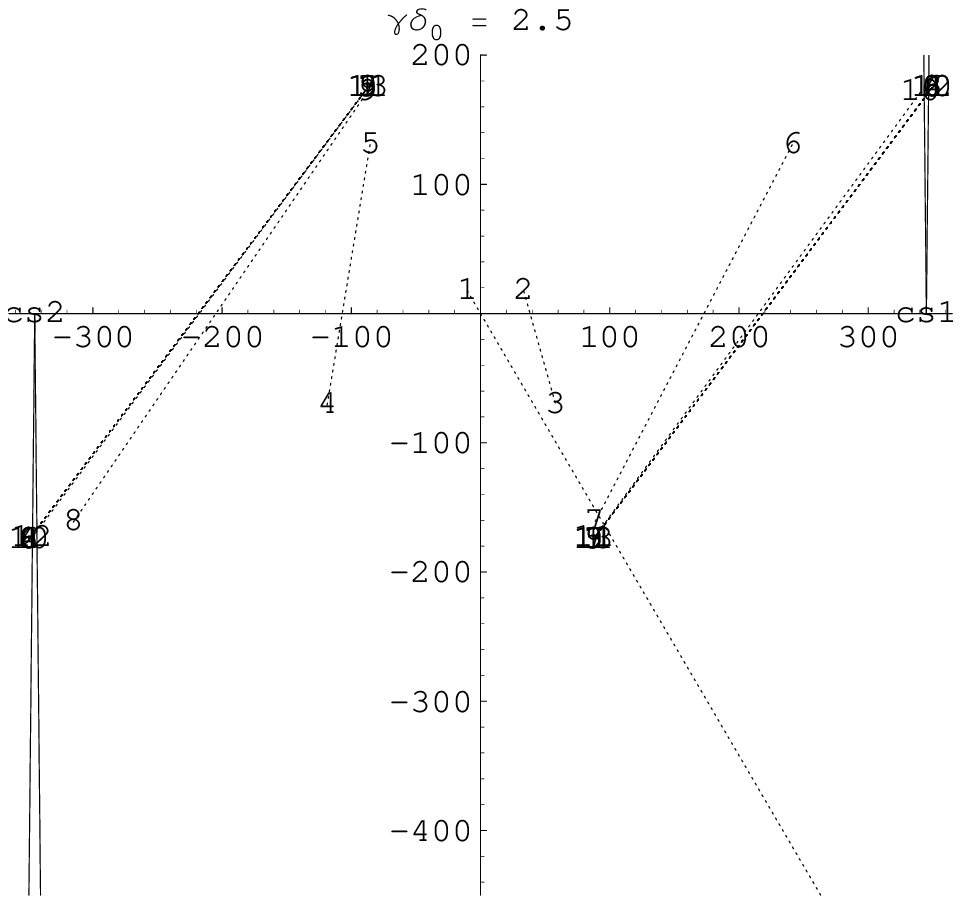,width=12cm}\\
\end{center}
\noindent {\bf FIG. 9}. The supercritical case.  Cosmic string \#1 (cs1) is moving to the left and
cosmic string \#2 is moving to the right.  The particle enters from the bottom right and
reaches cs1's deficit angle at $1$.  Its world line appears to jump to the
right to $2$, now directed down and slightly to the right. This process continues indefinitely, 
$3$, $4$, ...$\infty$, the world line spiraling clockwise outward as it falls 
into a stable orbit (the two long, outermost parallel segments).
The coordinate discontinuity seen here does not reflect any actual discontinuity. The numbers
on the world line satisfy $2n + 1 \equiv 2n + 2$ where the equivalence is identification via
Eq.(2).  The coordinate discontinuities plotted here are calculated with Eq.(\ref{xt3}), and the
trajectory angle is determined by Eq.(\ref{deltaalp}).  The initial conditions of the above trajectory
are not tuned, but rather generic.  Approximately half of all initial particle trajectories 
will end up in a CTC.

\vspace{4mm}

\begin{itemize}
\item photon momentum in the plane perpendicular to the strings will be blue-shifted 
as given by Eq.(\ref{blueshift}).  This formula applies to any relativistic particle.  
The blue shift occurs twice for each revolution, once from each string;

\item since momenta along the string lengths are not blue-shifted, non-relativistic 
particles and particles with velocities along the length of
the cosmic strings will be attracted to relativistic trajectories perpendicular
to the string length;

\item In the limit of small $d$ the average distance to the core of the CTC will not vary. This means
that the orbits will close, rather than shrink;

\item the energy of each particle in the CTC will diverge exponentially as a function
of number of cycles taken; since it takes no time to travel any number of cycles, infinite 
blue shift takes place;

\item therefore, $\gamma\delta_0 > 2$ implies a catastrophic divergence in the presence 
of even a single particle.

\end{itemize}

It should be noted that the exponential divergence in energy is kinematic, and has nothing
to do with particle number.
Below is pictured the trajectory of a photon in a supercritical Gott space.  The photon enters
from the lower right, and then spirals clockwise outward into a stable, blue-shifted orbit.
The upper and lower deficit angle wedges are moving to the left and right, respectively.

We thus
may conclude that although closed light-like curves may exist ($\gamma\delta_0 = 2$), a purely
classical divergence destroys the Gott solution
of closed time-like curves ($\gamma\delta_0 > 2$) in the presence of a dynamical field 
(e.g. gravitons or photons).  Cosmic string loops cannot produce closed time-like curves.
This is in agreement with Li and Gott \cite{Li:1997ka} which finds Misner space (and by implication 
Grant space and Gott space) to suffer from a classical instability similar to the one found here.
It should be noted that the hoop conjecture is not involved in this breakdown.
Even the 2+1 dimensional Gott space, where the hoop conjecture is not applicable, 
is unstable in the presence of dynamical fields.

\section{Comments}
%CTCs in quantum gravity}

\subsection{General discussions}

Semi-classical gravity raises an objection to the existence of CTCs, or at least to
spacetimes that contain both regions with CTCs and regions out of causal contact with
them.  The boundary between such regions is called the "chronology horizon," 
and in known examples this horizon coincides with a divergence of the 
renormalized energy-momentum tensor.  This led Hawking to pose the "chronology protection
conjecture," in which he proposes that any classical examples of spacetimes
containing CTCs will be excluded by a quantum theory of gravity.  
Thorne and Kim disagreed \cite{Kim:1991mc}
on the grounds that the divergence of $<T_{\mu\nu}>_{ren}$ is so weak that 
a full quantum gravity will remedy the semi-classical pathology.  Grant found that the divergence on a
set of
polarized hypersurfaces is much larger than that on the chronology horizon.
  String theory can directly address the conjectures made using semi-classical arguments.  Recent 
papers have evoked an Enhan\c{c}on-like mechanism as a stringy method to forbid the 
formation of CTCs in some spacetimes \cite{Jarv:2002wu}.  On the other hand, a recent paper by
Svendsen and Johnson demonstrated the existence of a fully string theoretic 
background (exact to all orders in $\alpha'$) containing CTCs and a chronology 
horizon, namely the Taub-NUT spacetime.
It is not clear if $g_s$ corrections destroy this result once matter is 
introduced, but the empty background is an exact result.  This seems to provide a counterexample 
to the Chronology Protection conjecture.

It is easy to see that a cosmic string with only the lowest mode will start as a circle and 
collapse to a point. Before it reaches that limit, a black hole is formed. An elliptic or 
rectangular loop can be made to collapse to parallel, relativistic segments
without forming a black hole. Although we argue that potential CTCs will be disrupted by the 
presence of photons (or any other mode), it is also possible that the huge blue-shift will cause 
the formation of a black hole, resulting in a black hole with cosmic string loops 
protruding. In this case, the presence of CTCs inside the black hole is acceptable, since 
they are not observable.  More analysis is needed to fully address this issue.

\subsection{Specific Discussions on 2 +1 Dimensions}

A simpler scenario with CTCs was found by van Stockum \cite{stockum} and Deser, Jackiw
and 't Hooft (DJtH) \cite{Deser:1983tn} 
whereby a single stationary cosmic string is
given angular momentum about its axis.  The resulting background is given by
\be
ds^2 = -(dt + J d\theta)^2 + dz^2 + dr^2 + (1 - 4G\mu)^2r^2 d\theta^2.
\ee
It seems unlikely for such a cosmic string to exist in string theory (at least as fundamental
objects\cite{Mazur:1986gb}), since
the cosmic strings in superstring theory lack the internal degree of freedom "spin".
DJtH noticed an unusual feature of the Gott spacetime.  They
classified the energy momentum of Gott's solution in terms of the Lorentz 
transformations
encountered under parallel transport (PT) around the strings (holonomy).  It is 
well known that the PT transformations around a single cosmic string is 
rotation-like, i.e. can be expressed as
\be
\Xi = \Lambda_\beta R_{\delta_0}\Lambda_{-\beta}
\ee
where $\Lambda_\beta$ is a pure boost with rapidity $\beta$ and $R_{\delta_0}$ is a pure
rotation about the angle $\delta_0$.  
One may calculate the holonomy of a Gott pair via
\be
\Xi = \Lambda_\beta R_{\delta_0}\Lambda_{-\beta}\Lambda_{-\beta}R_{\delta_0}\Lambda_\beta,
\ee 
and it is found that $\Xi$ is boost-like, i.e. $\Xi = \Lambda_\beta\Lambda_\zeta\Lambda_{-\xi}$.
DJtH regarded the energy momentum of a Gott pair to be unphysical on the grounds that
its holonomy matches that of a tachyon (boost-like).  It is an unusual feature
of spacetimes that are not asymptotically flat that $T^{\mu\nu}$ can be space-like (tachyonic)
despite the fact that it is made up of terms that are time-like.

Headrick and Gott \cite{GottH} refuted this criticism by showing that the Gott pair geometry was quite
unlike the tachyon geometry both because a tachyon does not yield CTCs and because the 
holonomy definition of $T^{\mu\nu}$ was incomplete. 
Later, Carroll, Farhi, Guth and Olum (CFGO) \cite{Carroll:1994hz} and 't Hooft \cite{tHooft2} 
gave more convincing arguments against CTCs. CFGO and Gott and Headrick found that the PT 
transformation of a spinor distinguished between tachyon and Gott pair geometries.  A 
more complete description of geometry would include not just the PT transformation, but
the homotopy class of the PT transformation as well.  Equivalently, one should extend
$SO(2,1)$ to its universal covering group.

One may interpret the ``boost-like'' holonomy as boost identification, as was done 
by Grant \cite{Grant:1992kj}.  This makes the Gott spacetime akin to a 
generalized Misner space.  Grant was able to show that the Gott/Misner spacetime suffers
from large quantum mechanical divergences on an infinite family of polarized light-like hypersurfaces.  
This divergence is stronger than that at the chronology horizon \cite{Grant:1992kj}.

Regardless of whether the Gott spacetime is physical or not, it can be shown that the Gott
spacetime in 2+1 dimensions cannot evolve from cosmic strings initially at rest \cite{Carroll:1994hz}.  
This is quite different from the 3+1 dimensional case.

\subsection{The Instability in 3 +1 Dimensions}

Recent realization of the inflationary scenario in superstring theory strongly suggests that cosmic superstrings were indeed produced toward the end of the inflationary epoch. With this possibility,
the issue has renewed urgency. In the above discussions, we argue that the reasoning against 
the Gott spacetime in 2+1 dimensions does not apply to the 3+1 dimensional case. 
In short, the Gott spacetime is entirely possible in an ideal classical situation. However, we argue that
instability set in if there is a quanta/particle nearby. The particle will be attracted to the closed 
time-like curve and is infinitely blue-shifted instantly. Of course, back reaction takes place before this happens. This back reaction must disrupt the closed time-like curve, otherwise the  infinite 
blue-shift will not be prevented. In an ideal situation where there is no quanta nearby, one still
expects particles like (very soft) gravitons/photons can emerge due to quantum fluctuation. In fact, quantum fluctuation of the cosmic strings themselves as they move rapidly toward each other
will produce graviton radiations. One graviton, no matter how soft, is sufficient to cause the instability.
So we believe that the Gott spacetime is unstable under tiny perturbations and so cannot be
formed in any realistic situation.

\section{Conclusion}

Recent implementation of the inflationary scenario into superstring theory led to the possibility that 
cosmic strings were produced toward the end of brane inflation in the brane world. This possibility 
leads us to re-examine the Gott spacetime, where closed time-like curves appear as two cosmic 
strings move ultra-relativistically pass each other. In an ideal situation, the Gott spacetime is an exact solution to Einstein equation, with a well-defined chronology horizon. It does not collapse into a black hole and can be readily reached. So it seems perfectly sensible that such a spacetime can be present 
in a universe that contains cosmic strings. In this paper, we find that nearby photons and gravitons will be attracted to the closed time-like curves, resulting in an instantaneous infinite blue-shift. This 
implies back-reaction must be large enough to disrupt the formation of such closed time-like curves. 
We interpret this as a realization of the chronology protection conjecture in the case of the 
Gott spacetime. 

\vspace{4mm}

{\large {\bf Acknowledgment}}

\vspace{2mm}

{We thank Robert Bradenberger, Eanna Flanagan, Gerard 't Hooft, Mark Jackson, Ken Olum, 
Geoff Potvin, Alex Vilenkin, Ira Wasserman, and Mark Wyman for discussions.
This work is supported by the National Science Foundation under Grant No. PHY-009831.}

\appendix

\section{Lensing by A Moving Cosmic String}

In the text, the lensing due to a moving cosmic string is too general for normal application.
Here we reproduce a simple elegant argument due to Vilenkin\cite{Vilenkin:1986cb} to calculate 
the angular separation of images produced by a moving cosmic string in the case when the 
cosmic string tension as well as the lensing effect are small.  We point out where the (small) error in Ref.\cite{Vilenkin:1986cb} is.
Vilenkin argues that the angular deflection of light by a cosmic string may be calculated
by appealing to Lorentz invariance.  For a string at rest, the angular separation is given by
\be
\delta\varphi  =  \frac{D_{s,cs}}{D_{s,O}}\delta_0,
\ee
where for simplicity we have assumed the string lies orthogonal to the line of sight.  We may consider
two light waves, one from each image, $k$ and $k'$.  Their scalar product is given by
\be
k^\mu k'_\mu = \omega\omega'(1 - \cos(\delta\varphi)) \approx \hf \omega\omega'(\delta\varphi)^2.
\ee
We may assume that the two light waves have the same frequency: $\omega = \omega'$. 
(This can be true in all reference
frames since we are expanding to first order in $\delta_0$.)  We can then relate the angular separation
in any two reference frames by the frequency of the light waves:
\be
\omega_0 \delta\varphi_0 = \omega\delta\varphi
\ee
i.e., the higher the observed frequency, the lower the observed angular separation.
The frequency in a reference frame where the string moves at 
velocity ${\bf v}$ relative to the observer is given by
\be
\omega = \frac{\omega_0}{\gamma(1 + \hat{{\bf n}}\cdot {\bf v})}
\ee
where $\hat{{\bf n}}$ is the direction from the observer to the source (and thus string), and hence
\be
\delta\varphi = \gamma(1 + \hat{{\bf n}}\cdot{\bf v})\delta\varphi_0.
\ee
The roles of $\omega$ and $\omega_0$ are erroneously swapped in Ref.\cite{Vilenkin:1986cb}, which thus agrees
with ours only for $\hat{{\bf n}}\propto{\bf v}$.
Physically, a traveler moving transverse to the light we observe should
measure a higher frequency than we do (i.e. $\omega < \omega_0$ for $\hat{{\bf n}}\cdot{\bf v}=0$).
It should also be pointed out that a string moving across the sky will blueshift the CMB behind it by an amount
\be
\frac{\omega_{back}}{\omega_{front}} = 1 + |{\bf v}\times\hat{{\bf n}}|\gamma\delta_0,
\ee
that is, the sky becomes hotter after a cosmic string passes.
\section{Branonium}

The analysis of relativistic bound states of D-Branes has been found to be
quite simple
in the probe brane
approximation (no radiation) \cite{Burgess:2003qv}.  We extend this to the
case of co-dimension
two branes (cosmic strings) using a similar analysis.
The action for a co-dimension two projectile is written as a
DBI-like term and a Wess-Zumino like term as follows:
\be
S = -\mu\int\sqrt{g} - q\int C_2
\ee
with
\be
C_2 = \log(r/r_0)dz\wedge dt 
\ee
We can now write down the canonical momenta associated to $\theta$ and $r$,
as well as the conserved Hamiltonian.  They are, respectively
\baray
\ell &=& \frac{r^2\dot{\theta}}{\sqrt{1-v^2}} \nonumber  \\
p_r &=& \frac{\dot{r}}{\sqrt{1-v^2}}  \nonumber   \\
H &=& \mu\sqrt{1 + p_r^2 + \frac{\ell^2}{r^2}} + q \log(\frac{r}{r_0})
\earay
where
\baray
v^2 &=& \dot{r}^2 + r^2\dot{\theta}^2  \nonumber  \\
\Rightarrow \frac{1}{\sqrt{1-v^2}} &=& \sqrt{1 +p_r^2 + l^2/r^2}.
\earay
The effective potential energy is plotted below.  Notable features are
the existence of stable orbits for generic initial conditions, and
exponentially long orbital periods as a function of energy.

Notice that the total energy of the system contains an arbitrary additive constant.  
This should be fixed by
the thickness of the strings, but it will have no effect on the dynamics. We
assume that the tension of the strings dominates the potential energy of the
long range interaction.
In this approximation, the geometry is flat except for a conical singularity at the 
location of each string.  For simplicity, we define the additive constant using the
perihelion distance (making it trajectory dependent): $r_0 = d$.  Then we may relate
the energy with the angular momentum as
\be
\epsilon = \sqrt{1 + \frac{\ell^2}{d^2}}.
\ee
A closed form solution for any trajectory can be found by making the
following substitution:
\baray
u &:=& 1/r \quad u' := \frac{du}{d\theta}  \nonumber  \\
\to u' &=& -\frac{p_r}{\mu\ell} \quad = \quad \frac{1}{\ell}\sqrt{(\frac{q}{\mu}
\log(ud) + \epsilon)^2 - 1 - u^2 \ell^2}
\earay
where we have used
\be
\frac{dr}{d\theta} = \frac{r^2p_r}{\mu\ell}  \nonumber 
\ee
and
\be
H = \mu\sqrt{1 + \ell^2(u'^2 + u^2)} - q\log(ud) = \mu\epsilon.  \nonumber 
\ee
Now we may solve for $\theta$ via
\baray
\theta = \int d\theta = \ell\int_{1/r_i}^{1/r_f}\frac{du}{\sqrt{(\frac{q}{\mu}\log(ud)
 + \epsilon)^2 - 1 - u^2\ell^2}}.
\earay

%%% \bibliography{\texdir Strings}

%\section*{References}

\end{document}